\documentclass[twocolumn,aps,prb,showpacs,floatfix]{revtex4}
\usepackage{graphicx}
\begin{document}
%%%%\draft

\title{Diffusion and jump-length distribution in liquid and amorphous Cu$_{33}$Zr$_{67}$}
\author{M. Kluge and H.~R.~Schober}
\affiliation{Institut f\"ur Festk\"orperforschung, Forschungszentrum
J\"ulich, D-52425, Germany}

\date{\today}
\begin{abstract}

Using molecular dynamics simulation, we calculate the distribution of atomic jumps
in Cu$_{33}$Zr$_{67}$ in the liquid and glassy states. In both states the
distribution of jump lengths can be described by a temperature independent
exponential of the length and an effective activation energy plus a
contribution of elastic displacements at short distances.
Upon cooling the contribution of shorter jumps dominates. No indication of
an enhanced probability to jump over a nearest neighbor distance 
was found. We find a smooth transition from flow in the liquid to jumps in the glass.
The correlation factor of the diffusion constant decreases
with decreasing temperature, causing a drop of diffusion
below the Arrhenius value, despite an  apparent Arrhenius law for
the jump probability. 
 
\end{abstract}
\maketitle

%\begin{multicols}{2}

\section{Introduction}
Metallic glasses are of great interest both fundamentally and due
to numerous applications. The absence of strong covalent bonds makes them 
the prime example of random dense packing.
Atomic transport in metallic glasses and their melts is generally considered
to be effected by two distinct mechanisms: flow in the liquid,
and thermally activated hopping in the glassy state. See Ref.~[\onlinecite{RMP}]
for a review. When a glass forming liquid is cooled
towards the glass transition diffusion drops faster
than predicted by an Arrhenius law. The diffusion coefficient
is often described by a Vogel-Fulcher law which can be rationalized e. g.
by a  free volume description.
Mode coupling theory
predicts for fragile glass formers, to which amorphous metallic
alloys belong, a dynamical transition at a critical temperature, $T_c$,
well above the glass transition temperature $T_g$. At $T_c$ flow
motion freezes and diffusion vanishes with a power law $D \propto
(T - T_c)^\gamma$ apart from a residual contribution from 
atomic hopping\cite{gotze:92}. The mechanism of this change from
flow to hopping is not understood.
 
In liquids with loose packing, flow is governed   
by binary collisions between the constituent atoms. In the undercooled
melt, where the atomic packing is dense, flow is strongly collective.
This is reflected in the observed small isotope effect of diffusion
\cite{ehmler:98}, similar to the one  observed earlier
in the glassy state \cite{faupel:90}. From these experiments it has been
concluded that, both below and above the glass transition, diffusion
is by collective motion of ten or more atoms. This poses the question of
whether the change from flow to hopping is a change to a new
elementary process or whether hopping evolves out of the flow motion.  

Over the last few years computer simulations have provided considerable
insight into the atomic dynamics of glasses and undercooled liquids.
Early molecular dynamics simulations (MD)
have shown collective jumps in undercooled liquids
\cite{miyagawa:88,wahnstrom:91}. Chains of atoms
replacing each other were observed, i.e. the single atoms jumped by a nearest 
neighbor distance. These chains can close to form rings 
\cite{miyagawa:88,teichler:97}.
Comparing successive configurations, averaged over typical vibrational times,  
one again finds chain-like structures of
atoms which have moved collectively in the undercooled liquid
\cite{SGO:97,donati:98,teichler:01}. These are not
necessarily replacement chains. 
The jump process in the glassy state  has been found
to involve many atoms, each single atom moving
only a fraction of the nearest neighbor distance in such a jump 
\cite{SOL:93,frank:94,ee:96}. 

Upon cooling towards the glass-transition a striking feature is
seen in the self-part of the van Hove function $G^s(r,t)$
which is related to the probability that an atom has moved by a distance
$r$ during a time $t$.
%\begin{equation}
%P(r,t)=4\pi r^2 G^s(r,t) = \langle \delta\left(r-\left|{\bf R}^n(t) -
%{\bf R}^n(0)\right| \right)\rangle .
%\end{equation}
At high temperatures $G^s(r,t)$ is perfectly Gaussian and broadens 
$\propto \sqrt{t}$. Upon cooling towards $T_c$, and beyond, a tail
to larger distances grows with time. Finally approaching $T_c$ an
additional second
peak at the nearest neighbor distance evolves, particularly for the more
mobile components. This effect can be taken as one of the signatures of
the glass transition \cite{roux:89}. From this behavior it was
concluded that there is a single peaked distribution of hopping
distances \cite{schroder:98}. 
The time evolution of $G^s(r,t)$
in CuZr could be reproduced by a simple model
involving jumps over nearest neighbor distances plus a residual small flow
\cite{GS:98}. In this picture the jump motion dominates the
diffusion in the undercooled liquid and the super-Arrhenius drop of 
diffusion stems from an increase of the return jump probability, as one
would expect from an increasing number of blocked paths.

In a quantitative investigation of the deviation of $G^s(r,t)$ from
a Gaussian, the non-Gaussianity parameter, $\alpha_2(T,t)$, was found to
increase rapidly in the undercooled liquid but no abrupt change near $T_c$ was
seen \cite{CMS:00,vollmayr:02}. The time evolution could be understood from a 
model of collective jumps \cite{CMS:00}.

Inspecting the pressure derivative of the diffusion constant, the apparent 
activation 
volume, one finds a strong cusp at $T_c$ which could indicate a change
of diffusion mechanism \cite{S:02}.  

These different findings pose the question of whether there is a change in
the elementary process of diffusion near $T_c$, in particular whether 
one might observe the evolution of a typical jump process in the glass.

\section{Simulation details}
Here we report a MD investigation of the atomic jump lengths in undercooled
and glassy Zr$_{67}$Cu$_{33}$.
For the inter-atomic interaction we use 
a modified Embedded Atom Method.\cite{baskes:94} 
The parameters were fitted to reproduce the experimental values of Cu, Zr
and CuZr$_2$ crystals. The universal energy-volume relation of 
Rose {\it et al.}\cite{rose:84} was used to determine the anharmonic
contributions, not sampled in the crystal but of essential importance in the 
disordered glassy state. We get lattice parameters $a=0.363$, $a=0.323,/,
c=0.516$ and $a=0.338, c=10.35$~nm (experimental values
\cite{kneller:86} $a=0.362$, $a=0.323$,
$c=0.515$ and $a=0.322$, $c=11.18$~nm) for Cu, Zr and CuZr$_2$, respectively.
The CuZr$_2$ lattice is slightly distorted. The atomic volume,
however, is only 2\% too large. The sublimation energies for Cu and Zr
(3.53 and 6.34~eV)
agree with experiment. We find enthalpies of fusion per atom relative to the
mono-atomic crystalline phases at room temperature of 
$\Delta_f H = 0.22$ and 0.18~eV 
(experiment \cite{zaitsev:03a} for CuZr$_2$ and CuZr, respectively.
The vacancy formation energies are $1.32$ and $1.63$~eV 
(experiment \cite{ehrhart:91} $1.28$ and $> 1.5$~eV) for Cu and Zr,
respectively. Additionally the phonon dispersion curves and elastic constants
of the mono-atomic lattices were used. In the case of Cu excellent
agreement was achieved. In Zr we get an overall agreement with 
experiment but some phonons 
deviate up to 30\%, similar to other work 
\cite{willaime:89,zhang:95}.
No attempt was made to fine tune the potential to fully reproduce 
the phase diagram.
The detailed form and the parameters are 
given in the appendix. For more details on the fitting 
procedure see   
Ref.~\cite{gaukel:thesis}. 

The MD calculations were done using the velocity Verlet algorithm with a 
time-step of $2.5 \cdot 10^{-15}$~s and systems of
$N=N^{\rm Zr}+N^{\rm Cu}= 1000$ atoms with periodic boundary conditions. 
Previous work on other systems (e. g. soft spheres, binary Lennard-Jones, Se)
by us and other groups
has shown that this size suffices to reproduce the dynamics at elevated
temperatures. As additional test some runs with $N=8000$ were done for
comparison. For the questions investigated in this work long aging times
are more important than large system sizes.
The pressure was kept constant
following Ref.~\cite{parrinello:80}
using a volume mass of $\approx \sqrt{N} \cdot m_{\rm Zr}$ and an additional
damping term to prevent oscillations.
Temperature was controlled by a Nos\'e-thermostat following 
Hoover \cite{hoover:85}.

\begin{figure}[htb]
\includegraphics*[bb=50 50 410 310,width=8cm,keepaspectratio]{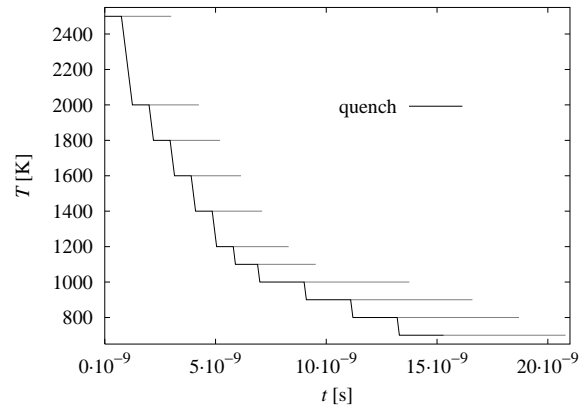}
\caption{Quench history of the samples (solid line). The dotted lines
show the lengths of additional aging before the start of the measuring 
runs.}
\label{fig_herstell}
\end{figure}

Three independent samples were prepared by a quench 
from the hot liquid and were aged in intermittent stages, 
as shown in Fig.~\ref{fig_herstell}.
We cool, in steps of 100~K, with a rate of 
$10^{12}$~K/s from 2000 K to the simulation temperature.  At each temperature 
step the samples
were aged for times ranging from 1~ns at 2000~K to 2~ns for 
$T \le 1000$~K, before
continuing the quench. The effective
cooling rate was thus lowered by about an order of magnitude, compared
to a straight quench with a constant rate. 
Before the actual
measurements at a given temperature the systems were additionally aged 
at constant $T$ for different times, up to 5.5 ns, as indicated by the 
dotted lines in Fig.~\ref{fig_herstell}.

%%%%%%%%%%%%%%%%%%%%%%%%%%%% fig.1 %%%%%%%%%%%%%%%%%%%%%%%%%%%%%%%%

\section{Glass transition and aging}

\begin{figure}[htb]
\includegraphics*[bb=50 50 410 310,width=8cm,keepaspectratio]{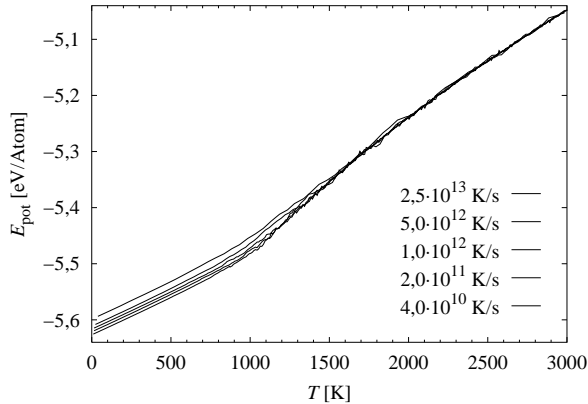}
\caption{Average potential energy per atom as function of temperature
for different quench rates: from top to bottom 
$Q = 2.5 \cdot 10^{13}, \; 5 \cdot 10^{12},  \; 1 \cdot 10^{12}, \; 
2 \cdot 10^{11}$ and $4 \cdot 10^{10}$~K/s.}
\label{fig_quench}
\end{figure}

Experimentally
${\rm Zr_{67}}{\rm Cu_{33}}$ is a good glass-former, which melts at
1310 K \cite{kneller:93} and can be under-cooled by a few hundred
 Kelvin. The experimental
glass-transition temperature varies from 600~K at cooling-rates of
0.01~K/s to 750~K at $10^6$~K/s \cite{kneller:93}.
To determine the glass transition temperature of our CuZr model
we did additional runs without the intermittent aging and
monitored the potential
energy and the volume as function of temperature for different quench rates, 
ranging
from $Q=2.5\cdot 10^{13}$~K/s to $Q=4\cdot 10^{10}$~K/s, 
see Fig.~\ref{fig_quench} for the potential energy. The glass transition
temperature, $T_{\rm g}$, was defined by the crossover from the low 
temperature to the
liquid behavior. In the limited range of $Q$ covered by the simulation
the dependence of $T_{\rm g}$ on $Q$ can be expressed by a
logarithmic law \cite{ritland:54} 
\begin{equation}
T_{\rm g} = \left( 463 + 20.6\cdot\ln{Q} \right)~{\rm K}.
\label{eq_Tg}
\end{equation} 
For our lowest quench rate $Q=4 \cdot 10^{10}$~K/s we find
$T_{\rm g} = 965$~K extrapolating to $Q=4 \cdot 10^{6}$~K/s 
Eq.~\ref{eq_Tg} gives 
$T_{\rm g} = 747$~K in excellent agreement with the experimental
value.

%%%%%%%%%%%%%%%%%%%%%%%%%%%% fig.2 %%%%%%%%%%%%%%%%%%%%%%%%%%%%%%%%

From Fig.~\ref{fig_quench} together with Eq.~\ref{eq_Tg} we find a linear
dependence of the average potential energy per atom  at $T=0$ on the
glass transition temperature
\begin{equation}
E_p(T=0,Q) = \left(-5.7715 + 1.5\cdot 10^{-4} T_{\rm g}(Q)\right)~{\rm eV}.
\label{eq_EofQ}
\end{equation}
Such a linear equation was also observed in amorphous Se for both energy
and atomic volume \cite{CS:02}. In CuZr the quench rate dependence of
the volume is too small to be evaluated.

Monitoring the average potential energy per atom at $T=1000$~K, 
near the glass transition temperature, starting at
the end of the quench,
Fig.~\ref{fig_herstell}, we find that it drops in 35~ns by about 0.014~ev.
The statistics is not sufficient to  determine the decay law unambiguously.

\section{diffusion and heterogeneity}

\begin{figure}[htb]
\includegraphics*[bb=30 60 550 660,width=7.5cm,keepaspectratio]{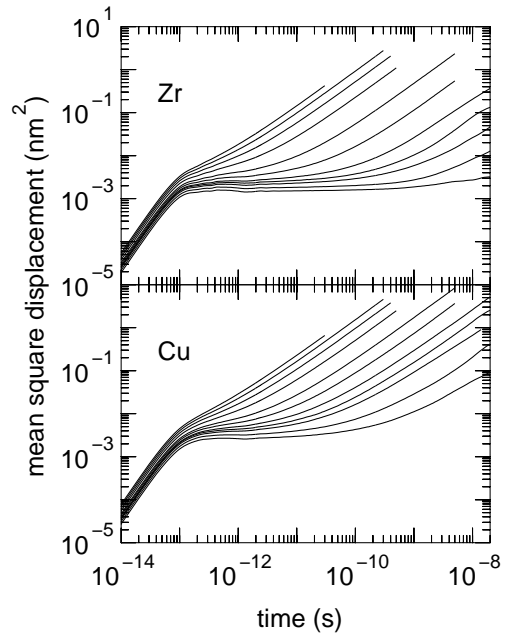}
\caption{Average mean square displacement as function of time on a
double logarithmic scale. Temperature from top to bottom: 2000, 1800,
1600, 1400, 1200, 1100, 1000, 960, 900, 800, 700~K. }
\label{fig_msqd_log}
\end{figure}

The average atomic mean square displacements (msqd), 
from the respective configurations after aging, are shown in 
Fig.~\ref{fig_msqd_log} in a double logarithmic plot. For short times one
observes an increase $\propto t^2$ which is typical for vibrational and
ballistic motion. For long times the msqd increases $\propto t$, indicative
of long range diffusion. Lowering the temperatures below $T=1400$~K the
plateau, typical for the undercooled liquid and the glass, evolves between 
these two limits. This onset of the plateau correllates well
with the experimental melting temperature
$T_m = 1310$~K \cite{kneller:93}.
At the lowest temperatures, in the ps range, one can just see some small
wiggles which reflect the vibration spectrum.  

\begin{figure}[htb]
\includegraphics*[bb=24 90 500 500,width=8cm,keepaspectratio]{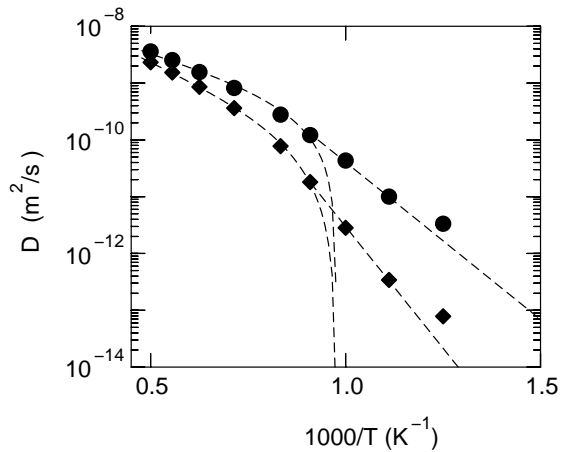}
\caption{Diffusion coefficients in ${\rm Zr_{67}}{\rm Cu_{33}}$ (Zr: diamonds,
Cu: spheres) The dashed lines represent a fit with MCT, using the same
temperature $T_c$ for both components, in the undercooled melt and an
Arrhenius fit in the glass.}
\label{fig_diffusion}
\end{figure}

The diffusion coefficients of the two components, Fig.~\ref{fig_diffusion}, 
were calculated in the usual way from the slope of
the long time limit of the msqd.
Fitting the diffusion coefficients in the undercooled melt
according to MCT,
\begin{equation}
D_{\rm MCT}(T) = D_0^{\rm MCT} / (T - T_c)^\gamma,
\label{eq_MCT}
\end{equation}
we find  $T_c = 1025$~K and $\gamma = 1.92$ and 1.34 for Zr and Cu, 
respectively. A fit to the relaxation time of the intermediate scattering
function with the same $T_c$ gives slightly higher values
$\gamma = 2.2$ and 1.57 for Zr and Cu, respectively \cite{kluge:thesis,RMP}. 
These numbers are
meant as a guide to the relevant temperatures and are not exact.
An alternative fit with the Volgel-Fulcher-Tammann (VFT) relation
\begin{equation}
D_{\rm VFT}(T) = D_0^{\rm VFT} \exp{(-E^{\rm VFT}/k(T-T^{\rm VFT}))}
\label{eq_VFT}
\end{equation}
gives for Zr $D_0^{\rm VFT} = 2.65 \cdot 10^{-8}$~m$^2$/s,
$E^{\rm VFT} = 0.28$~eV, $T^{\rm VFT} = 641$~K and for Cu
$D_0^{\rm VFT} = 3.70 \cdot 10^{-8}$~m$^2$/s,
$E^{\rm VFT} = 0.30$~eV, $T^{\rm VFT} = 477$~K.
In the glass the diffusivity can be described by the usual
Arrhenius law
\begin{equation}
D_{\rm Arrh}(T) = D_0^{\rm Arrh} \exp{(-E^{\rm m}/kT)}.
\label{eq_Arrh}
\end{equation}
Neglecting the values at $T=800$~K, which are probably too high due to
too short aging, we obtain
$D_0^{\rm Arrh} = 1.09 \cdot 10^{-3}$~m$^2$/s,
$E^{\rm m} = 1.7$~eV, for Zr and
$D_0^{\rm Arrh} = 1.41 \cdot 10^{-5}$~m$^2$/s,
$E^{\rm m} = 1.1$~eV, for Cu.
Due to the small fitting range there is a considerable margin of error on
these values. The deviation of the present values from the ones
reported earlier by Gaukel \cite{gaukel:thesis}
is due to the much improved statistics and aging of the present work. 
It should further be noted that fits in the undercooled liquid using
the VFT or MCT expression is strongly influenced by the range of temperatures
included in the fit. Furthermore in the fits of the previous work different 
values of $T_c$ were allowed for Cu and Zr, respectively, 
whereas in the present
work the condition of a unique value $T_c$ was imposed. The data do not suffice
to validate or invalidate this condition.

 Including the 800~K~values the activation energies would be considerably
smaller  
($E^{\rm m} = 0.99$ and  $0.75$~eV for Zr and Cu, respectively 
\cite{kluge:thesis,RMP}.
     
Our
results agree well with simulations of the similar NiZr system
where a totally different model for the inter-atomic interaction was 
used \cite{teichler:96}. The diffusion coefficients of the two components 
both in NiZr and 
in a binary Lennard-Jones \cite{S:02} glass at zero pressure are nearly
parallel in the melt, whereas in ${\rm Zr_{67}}{\rm Cu_{33}}$ they diverge.
This effect is probably due to the weaker
coupling between the two components in CuZr. Experimentally this is reflected 
in lower enthalpy of fusion in CuZr \cite{zaitsev:03a} compared to 
NiZr \cite{zaitsev:02}.

In isotropic diffusion the atomic displacements are Gaussian
distributed. In undercooled liquids and in glasses Gaussianity
is violated over long time scales. This non-Gaussianity
indicates different mobilities of different atoms over long 
time scales. This so called dynamic heterogeneity 
is quantified by the 
non-Gaussianity parameter \cite{rahman:64}
\begin{equation}
\label{eq_ngp}
\alpha_2(t)=\frac{3 <\Delta r^4(t)>}{5 <\Delta r^2(t)>^2}-1,
\label{eq_alpha}
\end{equation}
where $<...>$ denotes time averaging and $\Delta r^2(t)$ 
and $\Delta r^4(t)$ are the mean square and quartic displacements.

\begin{figure}
\includegraphics*[bb=70 100 520 440,width=8cm,keepaspectratio]{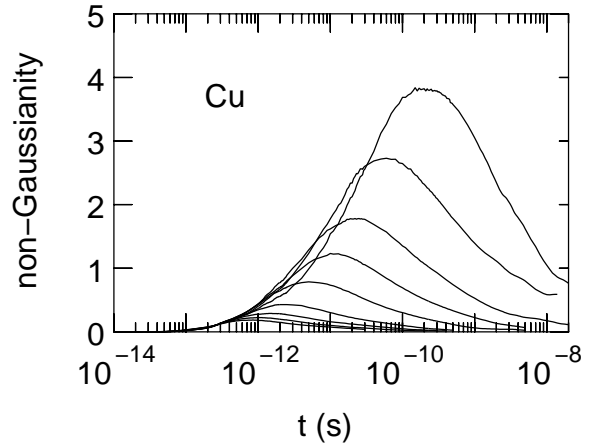}
\caption{Non-Gaussianity versus time of the Cu-subsystem in 
${\rm Zr_{67}}{\rm Cu_{33}}$ on a logarithmic scale. Temperatures from top
to bottom: 800, 900, 1000, 1100, 1200, 1400, 1600, 1800 and 2000~K.}
\label{fig_nongauss}
\end{figure}

Fig.~\ref{fig_nongauss} shows this non-Gaussianity in the Cu-subsystem
for temperatures from 800 to 2000~K. The general behavior resembles the
one observed in other glass formers. In the liquid above 1400~K dynamic
heterogeneity is weak and due to different local vibrational densities
of state. The maximal non-Gaussianity is at typical vibrational times (ps).
With increasing undercooling and even more in the glassy state 
the non-Gaussianity rapidly increases. The maximum is reached later and 
later. Comparing with Fig.~\ref{fig_msqd_log}, one sees that the decay
of the non-Gaussianity correlates with the onset of the diffusional part
of the msqd following the plateau. The increase of $\alpha_2(t)$ from
its vibrational value follows the $\sqrt{t}$ law which was attributed to
collective jump motion \cite{CMS:00}.

%%%%%%%%%%%%%%%%%%%%%%%%%%%% fig.2 %%%%%%%%%%%%%%%%%%%%%%%%%%%%%%%%

\section{Jump length distribution}

After aging we started the observation of jump processes. 
To get sufficient statistics we used 
observation times for the detection of jumps from  1.25~ns at 
1400~K to 2.5~ns below 1200~K. As definition of jump we 
use a rapid transit of an atom between two sites of residence. One can see
from a direct inspection of the particle trajectories that in
dense liquids or glasses such ``jumps'' are.
not ballistic. For the CuZr system 
such a trajectory
has been shown for a jump of a Cu-atom over a nearest neighbor distance in 
Refs.~[\onlinecite{kluge:thesis,GKS:99}].
Furthermore the amplitudes of short time excursions are very large which
makes an unambiguous definition of jumps difficult.

One method to detect atomic jumps is to study the ``inherent dynamics'' by
quenching the system at given time intervals ``instantaneously'' to 0~K 
and then study the properties of
the ``energy landscape''. Here we use a more direct approach and define jumps 
by differences between atomic
positions averaged over vibrational times. This means we define jumps of 
an atom in terms of absolute coordinates and not relative to the
neighbors of the atom.

As a first step we define for each
time step and each atom an average atomic position by
\begin{equation}
\langle{\bf R}^n(t)\rangle  = \frac{1}{\delta_1}
\int^{t+ 0.5\delta_1}_{t- 0.5\delta_1}
       {\bf  R}^n(t') dt' .
\label{eq_Rav}
\end{equation}
Since this averaging is done for each time step 
$\langle{\bf R}^n(t)\rangle$
is still, on the scale of the discretization by the simulation time step, 
a continuous function but is smoothed by averaging over the
vibrations. 
An instantaneous jump would cause a steep ramp in 
$\langle{\bf R}^n(t)\rangle$. 

For the detection of a jump we use two criteria. First, the
instantaneous position of an atom must differ by a minimum cutoff length
from the average taken at a previous time: 
\begin{equation}
\left| {\bf R}(t) - \langle{\bf R}^n(t-\frac{1}{2}\delta)\rangle \right| 
> r_1 .
\end{equation}
We want to exclude from our jump detection those excursions where
an  atom has a large amplitude momentarily, but immediately returns to its old 
site. Therefore, when the above condition is fulfilled we additionally
compare average positions
separated by the fixed time interval $\delta_2$
\begin{equation}
\Delta\langle{\bf R}^n(t)\rangle = \langle{\bf R}^n(t+0.5\delta_2)\rangle
                                    - \langle{\bf R}^n(t-0.5\delta_2)\rangle .
\label{eq_jump}
\end{equation}
A jump at time $t_0$ is recorded when this difference for the first time 
exceeds a limit
$r_2$
\begin{equation}
\left|{\Delta\langle{\bf R}^n(t_0)\rangle} \right| > r_2.
\label{eq_jump2}
\end{equation}
The corresponding atomic jump length is defined as
\begin{equation} 
\ell = \Delta\langle{\bf R}^n(t_0)\rangle .
\label{eq_ell}
\end{equation}

The time interval between the start time of the averaging for the final
configuration, $t+0.5\delta_2-0.5\delta_1$, and the end time of the 
averaging for the initial configuration,$t-0.5\delta_2+0.5\delta_1$
is
\begin{equation}
t_{\rm wait} = \delta_2 -  \delta_1.
\end{equation}

This method is rather CPU-time consuming since at each time step two
averages have to be done for each atom. This is however necessary if
one wants to get the necessary time resolution.

Throughout the simulation we used
the parameters $\delta_1 = 2.5$~ps, $\delta_2 = 4$~ps and $r_2 = 0.05$~nm.

The necessity of averaging the atomic configurations over given time intervals
limits the accuracy of the method.
If a jump is completed in the time interval, $t_{\rm wait}$, its length will 
be measured correctly by Eq.~\ref{eq_jump}. A longer ``jump time'' leads
to a reduced apparent jump length. On the other hand a movement of the
atom with constant velocity $v$ during the time $\delta_1 + \delta_2$
would be detected as jump if $v \cdot ( \delta_1 + \delta_2 ) > r_2$. 
Since averaging over typical vibrational times implies 
$\delta_1 \sim {\rm ps}$ we do not
expect this to be an important limitation.

More serious is the limited resolution of rapid successive jumps. If two
successive jumps are completed within the time interval $t_{\rm wait}$ they 
will
be regarded as a single jump. If, however, the second jump happens near
the end of the time interval it will be counted only partially. For example,
should 
the second jump be the reverse of the first jump we might under some
circumstances still record some shorter range jump, given by the sum of
the forward jump and a fraction of the back-jump. On the other hand, if
the first jump is rapidly followed by a successive forward jump, normally
correlated with the first jump, we record an effective jump length which
is too short. These two effects should approximately cancel for the 
investigated temperatures. For lower temperatures where the fraction
of back-jumps increases it leads to an overestimate of the mean square
displacement calculated from the jumps, compared to the exact value.   

Once the jump criteria hold for a time step they will normally
hold for a subsequent time interval. 
To avoid double counting, therefore after each jump,  we introduce 
a dead time $\delta_2$
during which jumps are not counted. This will lead to small inaccuracies,
mainly for the  shortest jump lengths.  
\begin{figure}[htb]
\includegraphics*[bb=30 100 550 615,width=8cm,keepaspectratio]{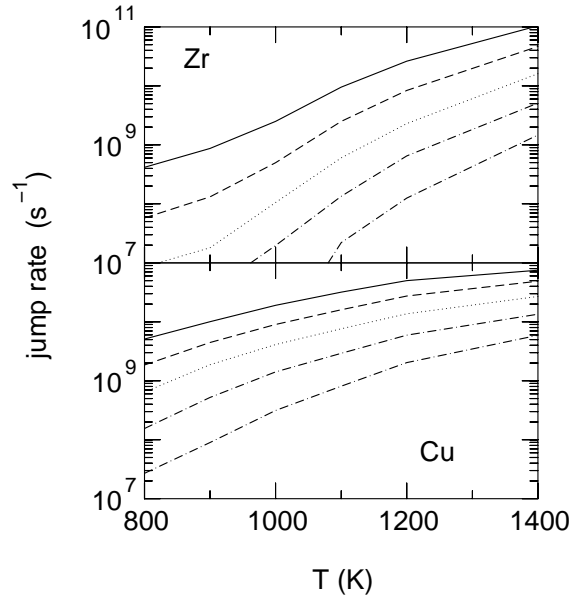}
\caption{Atomic jump rates, $ \gamma_{\rm jump}^j(\ell_{\rm cut})$,
on a logarithmic scale over temperature for Zr
for different cutoff lengths, $\ell_{\rm Cut}$: Zr (top) and Cu 
(bottom).
Lines from top to bottom: $\ell_{\rm cut} =0.06, 0.1, 0.15, 0.2,
0.25$~nm.}
\label{fig_jump_rates}
\end{figure}

From the recorded jumps we can calculate how often and how far each atom 
jumped during the observation time. This defines the average atomic jump rates
\begin{equation}
\gamma_{\rm jump}^j(\ell_{\rm cut}) =  \frac{1}{N^j t_{\rm obs}} 
\sum_{\ell < \ell_{\rm cut}} N^j_{\rm jump}(T,\ell,t_{\rm obs})
\label{eq_rate}
\end{equation}
where $N^j_{\rm jump}(T,\ell,t_{\rm obs})$ is the number of jumps of atoms of
species $j$ with
jump length in the interval $[\ell-\delta\ell /2,\ell+\delta\ell /2]$ with
$t_{\rm obs}$ and $T$ the observation time and temperature, respectively.
Fig.~\ref{fig_jump_rates} shows the rates for both components 
versus temperature for different cutoff 
length, $ \ell \le \ell_{\rm cut}$. As to be expected the jump rates
increase with temperature. No obvious break is apparent at $T_{\rm c}$ or 
$T_{\rm g}$. For the lower temperatures the jump rate for Zr is clearly
much smaller than that for Cu, particularly for the larger $\ell_{\rm cut}$.
This is in accordance with the lower diffusivity of Zr. Jumps over
nearest neighbor distances are observable for Zr only well above $T_{\rm g}$.
For Cu they are observed at all temperatures down to 800~K. At the highest 
temperature the rate for all jumps with $\ell \le \ell_{\rm cut} =0.06$~nm 
approaches
the theoretical limit of resolution of this simulation, 
$t_{\rm wait}^{-1} = 2.5 \cdot 10^{11}$~s$^{-1}$. This limits the applicability
of the method to higher temperatures. It reflects the gradual transition
from jumps to flow. It becomes meaningless to assign average starting
positions to the atoms. In Fig.~\ref{fig_jump_rates} average 
jump rates are shown. Individual
atoms will have higher rates. However, the dynamic heterogeneity is no
longer so important at $T = 1400$~K so that the deviations from the 
average are not too large. We will argue further down that the majority
of apparent jumps with short lengths is actually due to the elastic
displacement of surrounding atoms, accompanying all jumps.

For a more detailed investigation, we recorded for 
each detected jump the time of the jump and the initial and final
positions of the jumping atom, according to Eq.~\ref{eq_Rav}.
From these data the probability that an atom jumps by a certain
distance is calculated. Summing over the atomic jumps we also calculate
the atomic mean square displacements which can be compared with the ones
gained directly in the simulation.  

\begin{figure}[htb]
\includegraphics*[bb=30 100 550 615,width=8cm,keepaspectratio]{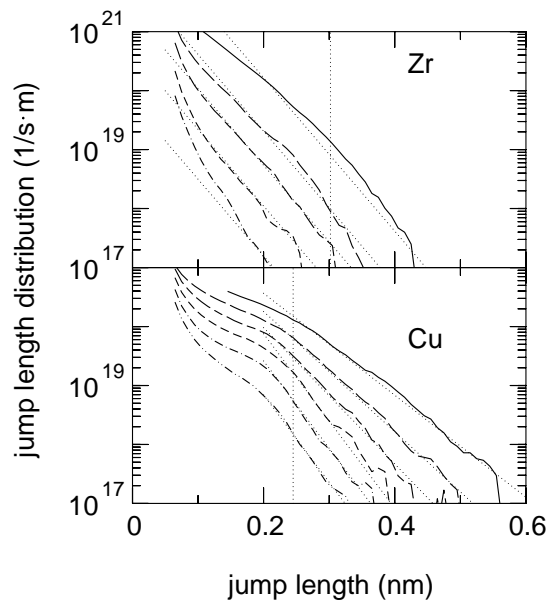}
\caption{Distribution of jumps/second over jump length for Zr (top) and Cu 
(bottom).
Temperatures from top to bottom: 1400, 1200, 1100, 1000 and 900~K. The 
dotted lines
indicate the fits by  exponential jump length distributions, see text. The 
respective nearest neighbor distances for the two components are indicated 
by the
vertical dotted lines.}
\label{fig_jump_distr}
\end{figure}

The probability that an atom jumps over a distance $\ell$
is  
\begin{equation}
P^j(T,\ell) \delta\ell = \frac{1}{N^j t_{\rm obs}} N^j_{\rm jump}(T,\ell,t_{\rm obs}).
\label{eq_length_distr}
\end{equation}
Fig.~\ref{fig_jump_distr} shows the distribution of the numbers of jumps 
per second
against jump length for the larger majority component Zr 
for temperatures ranging from 900 to 1400~K (top) and for the smaller
minority component Cu for temperatures ranging from 800 to 1400~K (bottom).
Looking first at the curves for Zr one clearly sees no indication of a 
preferred
jump length. The distribution can be fitted with a simple form
\begin{equation}
P_{\rm jump}(T,\ell) = A_{\rm jump} e^{-E_{\rm jump}/kT}e^{-\ell/\ell_{\rm jump}}
\label{eq_pjump}
\end{equation}
with $A_{\rm jump}^{\rm Zr} = 1.83 \cdot 10^{28}$~1/(m$\cdot$s), 
$E_{\rm jump}^{\rm Zr} = 1.51$~eV 
and $\ell_{\rm jump}^{\rm Zr} = 0.033$~nm.
The apparent activation energy $E_{\rm jump}^{\rm Zr}$ agrees within 
some 10\% with the
diffusional one in the glassy state. In the expression for the
undercooled melt, Eq.~\ref{eq_VFT}, it corresponds to the apparent activation
energy at $1140$~K.
As shown by the dotted line this fit works well in the whole temperature range
investigated which spreads over both $T_g$ and $T_c$. Of course $E_{\rm jump}^{\rm Zr}$ has to
be interpreted as an effective activation energy. There will be a spread of
activation energies which is absorbed by the prefactor $A_{\rm jump}^{\rm Zr}$.
The probability
of jumps over a nearest neighbor distance is two orders of magnitude less than
the one for jumps over half that distance.

In the case of Cu the situation seems more complicated. 
The probability
of jumps over a nearest neighbor distance is only one order of magnitude 
less than
the one for jumps over half that distance.
Different from Zr, there is a distinct tail to higher jump lengths whose 
origin is
not absolutely clear. One would expect a cutoff of 
$\Delta\langle{\bf R}^n(t)\rangle$
for distances not much greater than the nearest neighbor distance. In a 
densely
packed material a jump over larger 
distances is
highly improbable. We rather think that the tail is due to jumps which follow
so rapidly that we cannot resolve them. This view is supported also by
the correlation factors exceeding unity.
We will see below that the 
contribution of 
these ``extra long jumps'' drops at $T_c$ to less than 20\% of the 
total atomic 
mean square displacement.  

%There is 
%clearly no simple Arrhenius scaling of $P^{\rm Cu}(T,\ell)$
%in temperature. Different from Zr,
%on a logarithmic scale  the decay to large $\ell$-values becomes
%steeper with decreasing temperature. This would imply a larger
%``activation energy'' for long distance jumps. More likely, however,
%it just reflects the higher number of unresolved multiple jumps of
%the Cu-atoms at higher temperatures.  

%For short
%jumps, again a fit by a product of exponential dependencies on temperature 
%and 
%length is possible. The corresponding constants are
%$A_{\rm short jump}^{\rm Cu} = 2.35 \cdot 10^{23}$~1/sm, 
%$E_{\rm short jump}^{\rm Cu} = 0.53$~eV and 
%$\ell_{\rm short jump}^{\rm Cu1} = 0.083$~nm. 
%$E_{\rm short jump}^{\rm Cu1}$ is by a factor of 2 smaller 
%than the
%corresponding diffusional activation energy. However,
%this dependence applies only for a small
%section of the jump probability which gives only a minor contribution to
%the diffusion. Furthermore the range of applicability decreases with
%temperature.
For short jump distances we see again the ``elastic tail'' which evolves 
into a shoulder before it merges with the curve for the more important long 
jumps.
For these different fitting laws apply in the 
glass and in the undercooled liquid. In the glass, the Arrhenius form,
Eq.~\ref{eq_pjump}, works just as for Zr. The fitted values are:
$A_{\rm long jump}^{\rm Cu} = 9.79 \cdot 10^{26}$~1/(s$\cdot$m), 
$E_{\rm long jump}^{\rm Cu} = 0.82$~eV and 
$\ell_{\rm long jump}^{\rm Cu} = 0.024$~nm. This apparent activation energy is about
15\% less than the diffusional $E^m$ used in Fig.~\ref{fig_diffusion} but
is higher than the value gained after including $T=800$~K in the fit
of the diffusivity.

This fit breaks down in the undercooled liquid regime. The slopes of
the curves in Fig.~\ref{fig_jump_distr} decrease with temperature, at
variance with Eq.~\ref{eq_pjump}. Extending the fit from the glass into
the liquid one finds still quite good values for jumps over nearest 
neighbor distances. However for $\ell = 0.4$~nm at $T=1400$~K,  
Eq.~\ref{eq_pjump} already underestimates the jump probability by
nearly an order of magnitude. The jump length distribution for Cu in
the undercooled liquid can be fitted by an exponential law with a 
temperature dependent $\ell_{\rm long jump}(T)$
\begin{equation}
P_{\rm long jump}^{\rm Cu}(T,\ell) = B_{\rm long jump}^{\rm Cu} 
e^{-\ell/\ell_{\rm long jump}(T)} \;\;\; T>T_{\rm c}
\label{eq_pjump2}
\end{equation}
with $B_{\rm long jump}^{\rm Cu} = 1.96 \cdot 10^{22}$~1/sm and
$\ell_{\rm long jump}(T)= 0.038, 0.043$ and $0.05$~nm for $T= 1100, 1200$ 
and $1400$~K,
respectively. No direct physical origin for this relation is evident.
From the fact that the Arrhenius like temperature dependence, 
Eq.~\ref{eq_pjump}, is still applicable for jumps over nearest 
neighbor distances the most probable explanation seems to be that with
increasing temperature more and more rapidly succeeding jumps, which 
cannot be resolved into separate single jumps, occur with predominantly
additive direction. This is in accordance with the
increase in correlation factor discussed later on in this paper.

For both components we observe a rapid increase of the jump length distribution
toward short jump length. There is a clear excess above the value given by
Eq.~\ref{eq_pjump}. For Zr at $T=900$~K, where the feature is most prominent,
it can be fitted by $P_{\rm jump}(T,\ell) \propto 1/\ell^5$, close to the
$\propto 1/\ell^4$-dependence following from the $1/r$ decay of elastic 
displacements. We therefore conclude
that this rapid increase reflects the elastic displacements accompanying all
jumps. Due to dense packing they will also be present in the melt but are much
more prominent in the glassy state. Two effects will strongly reduce
their contribution to diffusion. First, the elastic displacement patterns are
different for each jump leading to a cancellation of the bulk of 
these displacements after a few jumps. In the present system this is 
clearly observed in the glass when the Cu-atoms are by more than an order
of magnitude more mobile than the Zr ones. The Cu-atoms move in a sluggish
matrix of Zr-atoms. Jumps of the Cu-atoms will be accompanied by 
displacements of the Zr-atoms without destroying their topology. After a
few subsequent
jumps, dominated by the faster Cu, the Zr will more or less be back to their 
original sites. Our algorithm will pick up the occasional large displacements
of the Zr but not their return in smaller steps. We will see this effect, 
further down, in the mean square
displacement of Zr at $T \le 900$~K, Fig.~\ref{fig_msqd_900}. 
temperatures the dynamic heterogeneity, meaning that at any given time only
a small subset of atoms is mobile, will increase this effect and spread it
to the second component. 
Secondly, after a jump process the glass is locally
excited  and relaxes toward the local equilibrium. This
is done by comparatively slight shifts of the average atomic positions.
This again reduces the contribution of the  small lengths to diffusion.

\section{Average mean square displacements}

\begin{figure}[htb]
\includegraphics*[bb=10 110 570 605,width=8cm,keepaspectratio]{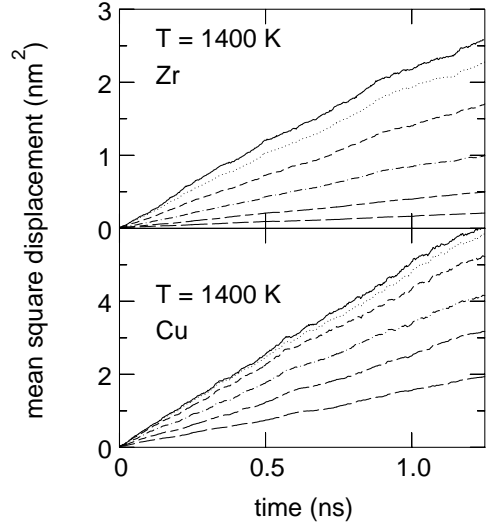}
\caption{Average atomic mean square displacement versus time at $T=1400$~K (full line). 
Atomic 
mean square displacement calculated from the jumps used for the  distribution, Fig.~1,
for different cutoffs: from top to bottom all jumps with jump lengths greater than
0.06, 0.1, 0.15, 0.2 and 0.25~nm.}
\label{fig_msqd_1400}
\end{figure}

\begin{figure}[htb]
\includegraphics*[bb=10 110 570 605,width=8cm,keepaspectratio]{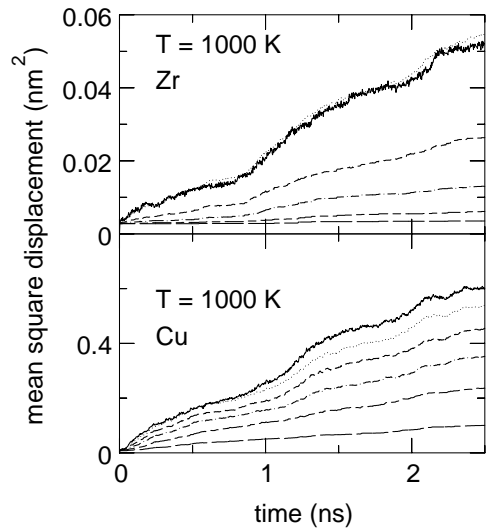}
\caption{Average atomic mean square displacement versus time at $T=1000$~K (full line). 
Atomic 
mean square displacement calculated from the jumps used for the  distribution, Fig.~1,
for different cutoffs: from top to bottom all jumps with jump lengths greater than
0.06, 0.1, 0.15, 0.2 and 0.25~nm.}
\label{fig_msqd_1000}
\end{figure}

To check the importance of the different jump lengths for diffusion we compare
the average atomic mean square displacement (msqd) taken directly from MD 
with the 
one obtained
by adding the jumps used to obtain the distribution of 
Fig.~\ref{fig_jump_distr}:
\begin{equation}
\left| \Delta R_{\rm jump}(t,\ell_{\rm cut})\right|^2 = 
\langle\ | \sum_{t'< t \atop \Delta{ R}^n > \ell_{\rm cut} } \Delta{\bf R}^n(t')\ |^2\
\rangle_n
\label{eq_msqd}
\end{equation}
where $\langle \dots \rangle_n$ indicates averaging over atoms and 
configurations.
We do this for different lower cutoffs of the jump length, $\ell_{\rm cut}$. 
In the undercooled melt, Figs.~\ref{fig_msqd_1400} and \ref{fig_msqd_1000},
there is an excellent agreement 
between the exact curves and the ones gained this way. In the figures, we added
to $\left| \Delta R^n_{\rm jump}(t,\ell_{\rm cut})\right|^2$ the vibrational 
msqd which can be obtained from the short time behavior.

We find for Zr that even at the temperature $T=1400$~K, i.e. about 40\% above
$T_c$, the msqd and, therefore also diffusion, is dominated by jumps much shorter
than the nearest neighbor distance, $R^{\rm Zr}_{NN} \approx 0.3$~nm. Jumps of
more than 0.25~nm contribute about 5\%. At $T=1000$~K their contribution vanishes.
In Cu, jumps over $R^{\rm Cu}_{NN} \approx 0.25$~nm give at $T=1400$~K about
a quarter of the msqd. At $T=1000$~K this contribution is diminished to 10\%.
This means that at $T=T_c$ short jumps dominate the diffusion of both 
components.   

\begin{figure}[htb]
\includegraphics*[bb=10 110 570 605,width=8cm,keepaspectratio]{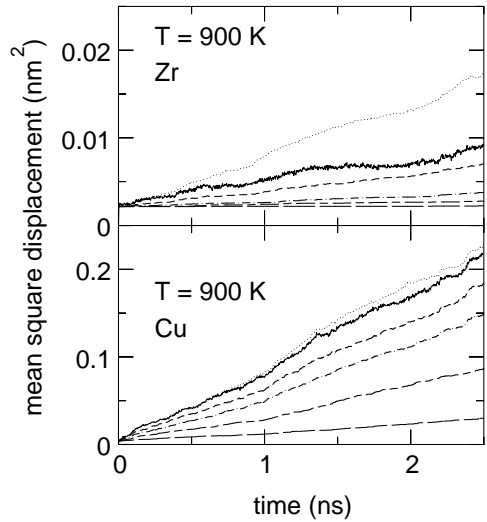}
\caption{Average atomic mean square displacement versus time at $T=900$~K (full line). 
Atomic 
mean square displacement calculated from the jumps used for the  distribution, Fig.~1,
for different cutoffs: from top to bottom all jumps with jump lengths greater than 0.06, 0.1, 0.15, 0.2 and 0.25~nm.}
\label{fig_msqd_900}
\end{figure}

At $T=900$~K, in the glass, the situation is similar for Cu.  For Zr the
contribution of long jumps is nearly negligible. The contribution of
short jumps, $\ell < 0.1$~nm and particularly $\ell < 0.06$~nm (dashed and
dotted lines in Fig.~\ref{fig_msqd_900}), is severely overestimated. This
can be traced to the ``elastic displacements'' discussed in the previous 
section. 
The motion of the Cu-atoms
in the Zr matrix causes displacements of the latter. Since the Zr subsystem
is sluggish compared with that of Cu, it has some memory over several
jumps of the Cu system. Zr-atoms will have a preference to return to their
original site. Or speaking in terms of potential energy, the system can
be approximated for some time as being in a minimum with the Cu motion as
perturbation. The Zr-atoms remain in their ``cages'' over many 
``cage-escapes'' of the Cu-atoms. While the Zr is in its cage it will
occasionally perform jumps picked up by the algorithm, followed by
several short jumps, not picked up. This can be seen by monitoring
the motion of the single atoms. An indication of this effect can be seen
in Fig.~\ref{fig_msqd_900} where, for
Zr, the dotted curve initially follows the real msqd (full line) but
later rises also when the real msqd of Zr stays more or less constant
while the msqd of Cu increases. This is the
discussed effect of Zr-atoms temporarily being displaced by Cu-jumps and
their return in several smaller steps not being detected. In this sense
the ``surplus'' of detected Zr-jumps corresponds to an in-cage motion.

We argued above that the rapid increase of $P^j(T,\ell)$ for $\ell \to 0$
is due to elastic displacements, i.e. displacements caused in the matrix
when an atom or a group of atoms jumps, and that these displacements will
not contribute strongly to diffusion. As diffusion drops
with sinking temperature, the ratio of probability of these non 
diffusive ``jumps''
over the one given by  Eq.~\ref{eq_pjump} rises strongly, see 
Fig.~\ref{fig_jump_rates}. If one does a simple correction of the curves 
for the two smaller cutoffs ($\ell_{\rm cutt} = 0.06$ and $0.1$~nm) by 
this ratio, the excess is removed and the msqd is actually underestimated
by 30\%. Such a simple correction does, of course, not distinguish 
between ``in cage'' and ``out of cage jumps''. These ``non-diffusive short
range jumps'' exist also in lattices. A jump of an atom into a neighboring
vacancy site causes displacements of the surrounding atoms which will
disappear again when the vacancy moves on. In lattices these can  
easily be measured and accounted for by means of lattice geometry and
symmetry. In an amorphous material this is no longer possible and it
becomes only a posteriori clear which of the short range displacements
contribute to diffusion and which are non-diffusive.    

\begin{figure}[htb]
\includegraphics*[bb=10 110 570 605,width=8cm,keepaspectratio]{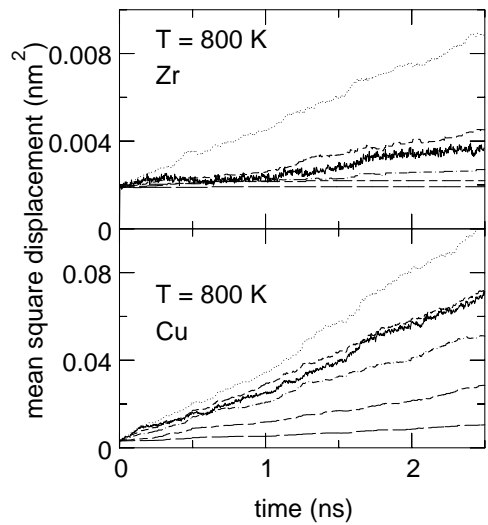}
\caption{Average atomic mean square displacement versus time at $T=800$~K (full line). 
Atomic 
mean square displacement calculated from the jumps used for the  distribution, Fig.~1,
for different cutoffs: from top to bottom all jumps with jump lengths greater than 0.06, 0.1, 0.15, 0.2 and 0.25~nm.}
\label{fig_msqd_800}
\end{figure}

Decreasing the temperature to 800~K the contribution of these elastic or
``in cage'' jumps increases further and becomes 
visible also for the more mobile Cu-atoms, Fig.~\ref{fig_msqd_800}.
This can be attributed to the
increasing dynamic heterogeneity. Like most of the Zr-atoms, an 
increasing number
of Cu-atoms becomes immobile on the time scale of several jumps. The
Cu-atoms, mobile at a given time, move in a matrix of Zr- and immobile
Cu-atoms which are temporarily displaced.

\section{Correlation factor}

In analogy to Eq.~\ref{eq_msqd} we can define an ``uncorrelated msqd'' by
\begin{equation}
\left| \Delta R_{\rm uncorr}(t,\ell_{\rm cut})\right|^2 = 
\langle \sum_{t'< t \atop \Delta{ R}^n > \ell_{\rm cut} } 
\left|\Delta{\bf R}^n(t') \right|^2\ \rangle_n
\label{eq_msqd_uncorr}
\end{equation}
and a correlation factor \cite{allnatt:87}
\begin{equation}
f_{\rm corr} = 
\lim_{t \to \infty} \frac{\left| \Delta R^n_{\rm jump}(t,\ell_{\rm cut})\right|^2}
{\left| \Delta R^n_{\rm uncorr}(t,\ell_{\rm cut})\right|^2 }. 
\end{equation}

Correlation factors have been studied in the past extensively for
diffusion in lattices \cite{flynn:72,philibert:91}. 
If the jumps are completely uncorrelated (random walk) one has 
$f_{\rm corr} = 1$. This would be the case for an isolated interstitial
atom, e.g. H in Pd, moving between equivalent sites. For diffusion by a
vacancy mechanism, neglecting lattice displacements, the correlation 
factor is reduced in simple lattices to
$ f_{\rm corr} \approx 1 - 2/z$ where $z$ is the number of nearest neighbors. 
The term $2/z$ accounts for the direct backjumps of the tracer atom into the
same vacancy. Exact calculations give for the vacancy mechanism in fcc
lattices $ f_{\rm corr} = 0.78$, whereas for a diamond lattice one has
$ f_{\rm corr} = 0.5$ due to the lower number of neighbors.
In simple lattices the correlation factor is given by geometry alone and is
temperature independent. In more complicated structures when more than
one type of jump is involved the correlation factor becomes temperature
dependent. An bias due to an external field increases the correlation 
factor.

If one would study self diffusion in a simple fcc lattice along the lines
of the two preceding sections one would obtain two sets of jumps with
two sets of correlation factors. ``Diffusive jumps'' would have a
jump length of about a nearest neighbor distance and a correlation factor
as discussed above. The short ``non-diffusive jumps'', i. e. the temporary
displacements due to the ``diffusive jumps'', on the other hand, would
have a correlation factor $ f_{\rm corr} = 0$. The total correlation
factor is thus somewhat reduced from the rigid lattice value.   
Total correlation factors zero are observed in lattices when atoms jump
between a finite number of sites, e. g. the
cage motion of interstitial Co in Al \cite{ERS:86}.

These general considerations assume that the diffusing particle is 
completely thermalized between jumps, i. e. that there are no memory
effects. At temperatures near the melting point this is no longer the
case. In simulations of vacancy diffusion in Al and Na lattices it was 
found, that when the
waiting time between successive jumps shrinks to the order of vibrational
time scales (ps), successive jumps become correlated, forward jumps
become more frequent and, subsequently, the correlation factor for vacancy
diffusion becomes larger than the random walk value, $ f_{\rm corr} > 1$ 
\cite{bennett:75,philibert:91}.

Transferring these results to the amorphous and undercooled liquid states,
one expects to find total correlation factors considerably smaller than 1
for diffusion via a vacancy type mechanism, and near to 1 for diffusion via
an inherent mechanism as postulated in the review \cite{RMP}. The correlation
factor for the longer jumps should be larger than for the shorter ones since
the latter have a larger non-diffusive contribution. Cooling below $T_g$,
the correlation factor should drop because more and more jump directions
become blocked. Unfortunately, due to the computational limitations, this 
effect cannot be explored fully at present. In the liquid, where we have seen
that the jump frequency becomes comparable to the vibration time, we 
expect an increase of the correlation factor with temperature. 

\begin{figure}[htb]
\includegraphics*[bb=10 60 570 605,width=8cm,keepaspectratio]{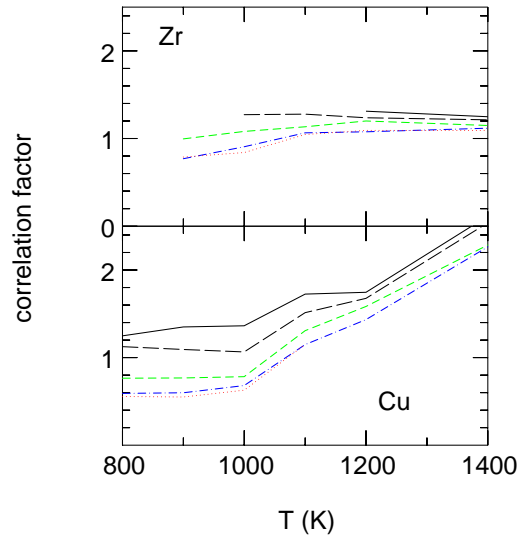}
\caption{Average atomic mean square displacement versus time at $T=800$~K (full line). 
Atomic 
mean square displacement calculated from the jumps used for the  distribution, Fig.~1,
for different cutoffs: from top to bottom all jumps with jump lengths greater than 0.25, 0.20, 0.15, 0.1 and 0.06~nm.}
\label{fig_correlation}
\end{figure}

The correlation factors calculated from the observed jumps are shown in
Fig.~\ref{fig_correlation}. By their very construction immediately following jumps are
excluded due to the dead time between jumps. The general trends are clearly seen in the
Fig. The correlation
factors for the long jumps are close to 1 or even larger, indicating 
diffusion by an inherent mechanism. The total correlation factor drops
upon cooling which explains the drop of the diffusion constant
below its Arrhenius line 
despite the seemingly constant effective activation
energy for the jumps themselves. Striking is the increase of the correlation
factor for Cu jumps to values around two for $T=1400$. This will be an
effect of insufficient thermalisation between jumps. It also indicates
the transition from jump to flow motion. For the larger and heavier 
Zr-atoms this effect is strongly suppressed.

\section{discussion}

Upon quenching, the mean square displacements, in a log-log representation, 
show for both components of Cu$_{33}$Zr$_{67}$
the normal behavior of a transition from a hot liquid, via
an undercooled one,  to a
glass. Between the $\propto t^2$ behavior, indicating ballistic motion or 
vibrations, and the $\propto t$ regime of diffusion a plateau develops.
The diffusion dynamics can be described by the
mode coupling scenario. However, the temperature dependences of the two
components do not run as parallel as in the often used binary Lennard-Jones
system either under constant volume \cite{kob:95a} or constant pressure 
conditions \cite{S:02}. Such a difference could be the origin of the order
of magnitude difference between diffusivity of Ni and Ti   and the one gained
via the Einstein relation from the viscosity
in ZrTiCuNiBe bulk glasses \cite{meyer:03} which was, similarly to the present
findings, explained by a faster
diffusion of the smaller components in a relatively rigid Zr-matrix. This behavior
seems to be typical for ZrCu systems. In a 
simulation of the binary NiZr 
system no such pronounced effect was found \cite{teichler:97}. 

Upon quenching, the dynamic heterogeneity rises rapidly above its 
vibrational value at temperatures below
$T=1400$~K which, in absence of a reliable estimate of the
melting point of our system, we take as a rough indication of undercooling.

The distribution of the jump lengths of both components, both above and below 
$T_{\rm c}$ is a smooth function of distance, definitely excluding a preference of 
nearest neighbor jumps.
Such a smooth distribution is to be expected for collective jumps as they
are seen in the typical chain-like motion of metallic glasses 
\cite{SOL:93,SGO:97,donati:98}. Using the energy landscape approach similar 
results were found for a binary Lennard-Jones system near $T_c$
\cite{schroder:00} and were also  attributed to cooperative motion which
in that system would be additionally furthered by the inherent higher
density. It is encouraging that the two different and complimentary approaches
lead to the same conclusion.

This result, of a missing typical jump length, will not be affected by the 
rapid quench inherent to simulations.
Both experiment \cite{RMP} and simulation of either activation volume
\cite{S:02} or isotope effect \cite{S:01} indicate that insufficient
aging would, if anything, enhance the number of single particle jumps with
their inherent length scale.

For the structure forming majority component Zr, the jump length distribution
is given by the product of two terms, an Arrhenius term for the 
temperature dependence and a temperature independent distribution of jump
lengths. An Arrhenius law for the temperature dependence, irrespective of 
the details of the barrier distribution has been derived
earlier in an effective medium treatment of hopping in disordered materials
\cite{wagener:91}. 
The spatial dependence shows a power law for short distances and and 
exponential one for the longer distances. This form is again in agreement
with the findings of the energy landscape study \cite{schroder:00}. The
power law part for short distances can be understood as being caused by the 
elastic 
displacement
accompanying any hopping process. These would be also seen for hopping in
a lattice where, different from a disordered system, they can easily 
identified from
translational symmetry. They are largely reverted by subsequent hops and 
will not 
contribute markedly to long range diffusion. The exponential long distance
dependence results, in our opinion, from the collectivity of the jumps. 
Jumps are closely correlated to low frequency 
quasi-localized vibrations \cite{SO:96}. The cores of these show, similarly
to the true localized vibrations, an exponential decay of the amplitudes
\cite{SR:04}. 

The jump  length distribution of the minority component, Cu,
has qualitatively the same behavior, however, with one important difference. 
The
Arrhenius scaling with temperature breaks down above $T_{\rm c}$. The
distribution is stretched to greater length and also the total number
of long jumps is increased. We have related this to lacking thermalisation
between jumps, a view supported by the correlation factors.

The correlation factors are of order unity. This is a
signature of diffusion by an inherent mechanism, as opposed to a defect
(quasi-vacancy) mediated one which would lead to clearly smaller correlation factors.
The correlation factor of Cu for $T>T_{\rm c}$ is an
exception. It increases strongly with temperature. Such correlation 
factors $>1$ are known to result when the time between jumps no longer
allows for thermalisation. It indicates, in this respect, 
the transition from jump-like motion
to flow.

From the contribution of the different jump length to the mean square
displacements it is obvious that long distance jumps only play a
minor role in Zr diffusion. Their role is greater for Cu where at $T_{\rm c}$
jumps over more than 0.2~nm contribute more than 30\%. From direct
inspection of individual jumps one sees, that in the usual chain of collectively
jumping atoms Cu-atoms
will show the largest displacements. At the higher temperatures, often more than
one Cu atom  jumps over a nearest neighbor distance. The replacement chains
observed in the early work \cite{miyagawa:88,wahnstrom:91}
are a special case of this scenario.

The absence of a preferred jump length might, prima facie, be taken as a
contradiction to the time evolution of the van-Hove self correlation
function where, in the undercooled liquid, clearly a secondary peak at
the nearest neighbor distance evolves. One of the authors showed earlier
for the same material 
that the time evolution of the 
van-Hove self correlation function for Cu near $T=T_{\rm c}$ can be 
reproduced by
a model comprising only nearest neighbor jumps and a residue of short
distance jumps, called flow-like motion \cite{GS:98}. For Cu an apparent
activation energy of 850~meV, in good agreement with the present value
of 820~meV, was given. Combining this and the present results one cane
conclude that the secondary peak in the van-Hove function indicates
preferred resting positions for the Cu-atoms in the Zr-matrix which is rigid
on the time scale of the Cu diffusion. 
A similar conclusion was arrived at for the Lennard-Jones system from
the energy landscape picture \cite{schroder:00,buchner:00} when the dynamic 
heterogeneity
creates a sufficiently rigid matrix of immobile particles.
Diffusion 
becomes dominated 
by transitions between these minima, a connected network of minima 
is formed \cite{stillinger:95,angelani:98,schulz:98} or
``coarse graining'' occurs \cite{rinn:01}.
The strongly different diffusivities of the present system enhance the effect and
are an additional source of ``metabasins'' \cite{doliwa:03}. 
The
diffusion of the Cu in the deeply undercooled liquid and even more in the glass has
features of a diffusion between ``traps''. 

In the previous work \cite{GS:98}, as the 
temperature is lowered, an increase in jump reversal was reported. 
This is compatible with the reported correlation
factors of the present work. 
A relative increase of reversible jumps for lower 
temperatures has also been observed from a jump analysis of the binary Lennard-Jones
system below $T_{\rm c}$ \cite{vollmayr:04}. 

\section{Summary}
In summary we calculated the diffusion of Cu and Zr above and below the
glass transition temperature. Different to the commonly studied Lennard-Jones
system Cu shows in the undercooled liquid a strongly enhanced diffusivity, 
compared
to the majority component Zr.
We think that this effect is related to the weaker coupling between the
two components in the present system which is seen also in the reduced
enthalpy of fusion.

The distribution of atomic jumps in the liquid and
glassy state can be described by simple exponential 
dependencies
on jump length and temperature. There is a smooth transition from flow to 
hopping. For the faster Cu this is reflected in the undercooled liquid by
correlation factors exceeding unity, indicating a breakdown of thermalisation
between jumps.
We find no preferred jump length around the
nearest neighbor distance. The observed secondary peaks in the van-Hove 
function
are, therefore, clearly not an effect of the jumps directly but of increased
waiting times at given sites. This effect will be enhanced by the growth
of the dynamical heterogeneity upon cooling. On the time scale
of the inverse of the jump frequency of the mobile atoms, 
more and more atoms are immobile
and provide a semi rigid background thus creating preferred sites. The effect
is seen more strongly by the smaller atoms (Cu). 

\appendix
\section{Modified embedded atom interaction (MEAM)}

We use the MEAM model developed for the CuZr system by Gaukel 
\cite{gaukel:thesis}. For completeness we give below the analytic form 
and the parameters.
  
In the Embedded Atom Method (EAM) model \cite{daw:83} the interatomic
interaction is described by pair potential $\Phi_{\rm two}$ and an 
embedding energy $F_{\rm ea}(\rho)$ which accounts for the additional
many body effects due to the electronic density:
\begin{equation}
  \label{equ:interaction_potential}
  E_{\rm pot}
  =
  \sum_{m,n=1 \atop m \not= n}^N \frac{1}{2} \Phi_{\rm two}^{\ell^m\ell^n} 
\left( R^{mn} \right)
  +
  \sum_{n=1}^N F_{\rm ea}^{\ell^n} \left( \rho \left( \textbf{R}^n \right) \right)
\end{equation}
where $R^{mn} = \left| \textbf{R}^m - \textbf{R}^n \right|$ is the distance
between atoms $m$ and $n$, and 
$\ell^n$ indicates the type of atom, in our case Zr or Cu.

For the pair potential we use the analytic form
\begin{eqnarray}
  \label{equ:interaction_potential_two}
  \Phi_{\rm two} \left( R^{mn} \right)
  &=&
  c_1 e^{ - c_2 R^{mn}} 
  + c_3 e^{- c_4 \left( R^{mn} - R_0 \right) ^2} \nonumber \\
  & + & c_5 \left( R^{mn} \right) ^6
   + c_6 \left( R^{mn} \right) ^7 
   + c_7 
\end{eqnarray}
The small parameters $c_5, c_6$ and $c_7$ are given by the condition
that $\Phi_{\rm two}$ and its first two derivatives vanish at $R_{\rm cutoff}$.
The parameters are given in Table~\ref{tab:pot_ener_two} for Zr and Cu.
For the mixed interaction we use the the mean
\begin{equation}
\Phi_{\rm two}^{\rm ZrCu}\left( R^{mn} \right) = \left(\Phi_{\rm two}^{\rm ZrZr}\left( R^{mn} \right) + \Phi_{\rm two}^{\rm CuCu}\left( R^{mn} \right)\right)
/2.
\end{equation}

\begin{table}[htb]
  \begin{center}
\begin{tabular}{|l||c|c|}
  \hline
  \rule[-0.7em]{0em}{1.9em} & Zr & Cu \\
  \hline
  \hline
  \rule[-0.7em]{0em}{1.9em} $c_1 \;\left[\mbox{eV}\right]$ & $3.457155  \cdot 10^4$ & $6.547955  \cdot 10^5$\\ 
  \hline
  \rule[-0.7em]{0em}{1.9em} $c_2$ & $4.479563$ & $6.487234$ \\ 
  \hline
  \rule[-0.7em]{0em}{1.9em} $c_3 \;\left[\mbox{eV}\right]$ & $-1.062312$ & $-2.383022$  \\ 
  \hline
  \rule[-0.7em]{0em}{1.9em} $c_4$ & $0.8614267$ & $0.1318588$ \\ 
  \hline
  \rule[-0.7em]{0em}{1.9em} $R_0 \;\left[\mbox{\AA}\right]$ & $2.952510$ & $0.6049550$ \\
  \hline
  \rule[-0.7em]{0em}{1.9em} $c_5 \;\left[\mbox{eV}/\mbox{\AA}^6\right]$ & $-4.971823 \cdot 10^{-8}$ & $-3.085439 \cdot 10^{-4}$ \\ 
  \hline
  \rule[-0.7em]{0em}{1.9em} $c_6 \;\left[\mbox{eV}/\mbox{\AA}^7\right]$ & $6.337047  \cdot 10^{-9}$ & $5.310256  \cdot 10^{-5}$ \\ 
  \hline
  \rule[-0.7em]{0em}{1.9em} $c_7 \;\left[\mbox{eV}\right]$ & $6.636436  \cdot 10^{-4}$ & $0.9001767$  \\
  \hline
  \rule[-0.7em]{0em}{1.9em} $R_{\rm cutoff} \;\left[\mbox{\AA}\right]$ & $6.57620462$ & $4.44761582$  \\
  \hline
\end{tabular}
\caption{The constants of the pair potential, Eq.~\ref{equ:interaction_potential_two}.}
    \label{tab:pot_ener_two}
  \end{center}
\end{table}

For the embedding term, we use the form proposed by 
Baskes \cite{baskes:87} for both components
\begin{equation}
  \label{equ:interaction_potential_ea}
  F_{\rm ea} \left( \rho \left( \textbf{R}^n \right) \right) = c_1 \rho
  \left( \textbf{R}^n \right) \cdot \ln ( c_2 \rho \left( \textbf{R}^n
  \right) )
\end{equation}
with the parameters given in Table~\ref{tab:pot_ener_ea}.
\begin{table}[htb]
  \begin{center}
    \leavevmode
\begin{tabular}{|l||c|c|}
  \hline
  \rule[-0.7em]{0em}{1.9em} & Zr & Cu \\
  \hline
  \hline
  \rule[-0.7em]{0em}{1.9em} $c_1 \;\left[\mbox{eV} \cdot \mbox{\AA}^3\right]$ & $2.71082977$  & $2.02891465$  \\ 
  \hline
  \rule[-0.7em]{0em}{1.9em} $c_2\;\left[\mbox{\AA}^3\right]$                  & $0.737740301$ & $0.876399217$ \\ 
  \hline
\end{tabular}
\caption{The constants used in the embedding term, Eq.~\ref{equ:interaction_potential_ea}.}
    \label{tab:pot_ener_ea}
\end{center}
\end{table}

In the original EAM, the density $\rho$ in 
Eq.~\ref{equ:interaction_potential} and Eq.~\ref{equ:interaction_potential_ea}
is given as a superposition
of radial functions 
\begin{equation}
  \label{equ:interaction_potential_density_eam}
  \rho_0 \left( \textbf{R}^n \right) = \sum_{m=1 \atop m \not= n}^N f(R^{mn}).
\end{equation}
For the function $f(R^{mn})$ we use an exponential plus additional terms to
set the function and its first two derivatives to zero at the cutoff.
\begin{equation}
  \label{equ:interaction_potential_density_0_analytic}
  f(R^{mn})
  =
   c_1 e^{- c_2 R^{mn}} 
  + c_3 \left( R^{mn} \right) ^{c_4}
  + c_5 \left( R^{mn} \right) ^{c_4+1} \nonumber
  + c_6 \nonumber 
\end{equation}

The parameters are compiled in Table.~\ref{tab:pot_ener_density_0}
\begin{table}[htb]
  \begin{center}
\begin{tabular}{|l||c|c|}
  \hline
  \rule[-0.7em]{0em}{1.9em} & Zr & Cu \\
  \hline
  \hline
  \rule[-0.7em]{0em}{1.9em} $c_1 \;\left[1/\mbox{\AA}^{3}\right]$ & $1.04537268$  & $0.896989894$  \\ 
  \hline
  \rule[-0.7em]{0em}{1.9em} $c_2$ & $0.123135389$ & $0.286315587$ \\ 
  \hline
  \rule[-0.7em]{0em}{1.9em} $c_3 \;\left[1/\mbox{\AA}^{3}\right]$ & $5.54697310 \cdot 10^{-12}$ & $6.74032288 \cdot 10^{-6}$ \\ 
  \hline
  \rule[-0.7em]{0em}{1.9em} $c_4$ & $15.1751328$ & $7.51090487$ \\ 
  \hline
  \rule[-0.7em]{0em}{1.9em} $c_5 \;\left[1/\mbox{\AA}^{3}\right]$ & $-9.43929033 \cdot 10^{-13}$ & $-1.22957801 \cdot 10^{-6}$ \\ 
  \hline
  \rule[-0.7em]{0em}{1.9em} $c_6 \;\left[1/\mbox{\AA}^{3}\right]$ & $-0.597896189$ & $-0.345100552$ \\ 
  \hline
  \rule[-0.7em]{0em}{1.9em} $R_{\rm cutoff} \;\left[\mbox{\AA}\right]$ & $5.16447906$ & $4.28405051$ \\
  \hline
\end{tabular}
\caption{Parameters of the spherical density function,
Eq.~\ref{equ:interaction_potential_density_0_analytic}.}
\label{tab:pot_ener_density_0}
\end{center}
\end{table}

In the MEAM additional angular terms are added to allow for covalent effects
\cite{baskes:87} needed to describe non ideal hcp lattices. We restrict
ourselves to terms in the third power of the cosine of the apex angle
\begin{displaymath}
  \cos (\Theta^{mnl}) 
  =
  \left(
    \frac{\textbf{R}^{mn}\,\textbf{R}^{ln}}{R^{mn}\,R^{ln}}
  \right)
  .
\end{displaymath}
The density then takes the form
\begin{eqnarray}
  \label{equ:interaction_potential_density}
  \rho \left( \textbf{R}^n \right)
  =
  \rho_0 \left( \textbf{R}^n \right) \exp \left[ \frac{1}{(\rho_0 \left( \textbf{R}^n \right))^2}  c_3  \right. \nonumber\\
  \left.   \times     \sum_{l,m=1 \atop {l \not= n \atop m \not= n} }^N
                       \cos^3 ( \Theta^{mnl} ) \;
                       f_3^{\ell^m}(R^{mn}) \; f_3^{\ell^l}(R^{ln})  \right] .
\end{eqnarray}

The angular correction in the MEAM is only needed for Zr as apex atom 
but not for Cu.
We, therefore, put $c_3 = 0$ for Cu, Table~\ref{tab:pot_ener_density}
\begin{table}[htb]
  \begin{center}
\begin{tabular}{|l||c|c|}
  \hline
  \rule[-0.7em]{0em}{1.9em} & Zr & Cu \\
  \hline
  \hline
  \rule[-0.7em]{0em}{1.9em} $c_3$ & $-5.36035659$ & $0$ \\ 
  \hline
\end{tabular}
\caption{Parameters of the angular correction in the MEAM, 
Eq.~\ref{equ:interaction_potential_density}.}
    \label{tab:pot_ener_density}
\end{center}
\end{table}

For the radial function in the angular correction term we used the same
form as for the spherical part
\begin{equation}
  \label{equ:interaction_potential_density_m_analytic}
  f_3(R^{mn})
  =
  c_1 e^{- c_2 R^{mn}}
  + c_3 \left( R^{mn} \right) ^{c_4} 
  + c_5 \left( R^{mn} \right) ^{c_4+1}
  + c_6
\end{equation}
with the parameters given in Table~\ref{tab:pot_ener_density_m}.
\begin{table}[htb]
  \begin{center}
\begin{tabular}{|l||c|c|}
  \hline
  \rule[-0.7em]{0em}{1.9em} & Zr & Cu \\
  \hline
  \hline
  \rule[-0.7em]{0em}{1.9em} $c_1 \;\left[1/\mbox{\AA}^{3}\right]$ & $1.04537268$ & $0.896989891$\\
  \hline
  \rule[-0.7em]{0em}{1.9em} $c_2$ & $0.146268549$ & $1.0000000$ \\
  \hline
  \rule[-0.7em]{0em}{1.9em} $c_3 \;\left[1/\mbox{\AA}^{3}\right]$ & $2.37968067 \cdot 10^{-14}$ & $2.67398161 \cdot 10^{-5}$ \\
  \hline
  \rule[-0.7em]{0em}{1.9em} $c_4$ & $19.5094567$ & $6.00000000$\\
  \hline
  \rule[-0.7em]{0em}{1.9em} $c_5 \;\left[1/\mbox{\AA}^{3}\right]$ & $-4.53181412 \cdot 10^{-15}$ & $-5.15696452 \cdot 10^{-6}$ \\
  \hline
  \rule[-0.7em]{0em}{1.9em} $c_6 \;\left[1/\mbox{\AA}^{3}\right]$ & $-0.57983725$ & $-4.14635228 \cdot 10^{-2}$ \\
  \hline
  \rule[-0.7em]{0em}{1.9em} $R_{\rm cutoff} \;\left[\mbox{\AA}\right]$ & $4.74779381$ & $4.0000000$\\
  \hline
\end{tabular}
\caption{Parameters of the angular density function, 
  Eq.~\ref{equ:interaction_potential_density_m_analytic}.}
    \label{tab:pot_ener_density_m}
\end{center}
\end{table}

%\bibliography{md}

%\end{multicols}
\end{document}